\documentclass[sigplan,screen,dvipsnames]{acmart}
\usepackage{xspace}
\usepackage{caption}
\usepackage{subcaption}
\usepackage{grumble}
\usepackage{listings}
\usepackage{tcolorbox}
\usepackage{longtable}
\usepackage{comment}
\usepackage{xurl}
\usepackage{subcaption}
\usepackage{booktabs}
\usepackage{tagging}
\usepackage{multirow}
\usepackage{colortbl}
\usepackage{tikz}
\usepackage{microtype}
\usetikzlibrary{quantikz}
\usepackage{algorithm}
\usepackage{algpseudocode}

\usepackage[nounderscore]{syntax}
\usepackage{color}
\usepackage{adjustbox}

\usepackage[inline]{enumitem}
\setlist{noitemsep,topsep=0pt,parsep=0pt,partopsep=0pt}

\usepackage[subtle,paragraphs=normal]{savetrees}

\usepackage{setspace}
\usepackage[margin=5pt,font={stretch=0.9}]{caption}

\AtBeginDocument{
  }

\setcopyright{cc}
\acmDOI{10.1145/3696443.3708965}
\acmYear{2025}
\setcctype{by} 
\copyrightyear{2025}
\acmISBN{979-8-4007-1275-3/25/03}
\acmConference[CGO '25]{Proceedings of the 23rd ACM/IEEE International Symposium on Code Generation and Optimization}{March 01--05, 2025}{Las Vegas, NV, USA}
\acmBooktitle{Proceedings of the 23rd ACM/IEEE International Symposium on Code Generation and Optimization (CGO '25), March 01--05, 2025, Las Vegas, NV, USA}
\received{2024-09-09}
\received[accepted]{2024-11-04}
\begin{document}

\title[\projectname: A Retargetable Compiler Framework for FPQA Quantum Architectures]{\projectname: A Retargetable Compiler Framework \protect\\ for FPQA Quantum Architectures}

\newcommand{\projectname}{\textsc{Weaver}\xspace}
\newcommand{\wqasm}{\textsc{wQasm}\xspace}
\newcommand{\wcompiler}{\textsc{wCompiler}\xspace}
\newcommand{\woptimizer}{\textsc{wOptimizer}\xspace}
\newcommand{\wchecker}{\textsc{wChecker}\xspace}

\newcommand{\myparagraph}[1]{\noindent{\bf {#1}.}}

\newcommand*\circled[1]{\tikz[baseline=(char.base)]{
            \node[shape=circle,draw,inner sep=1pt] (char) {#1};}}
            
\newcommand{\takeaway}[1]{
    \vspace{2mm}
    \noindent\fbox{\parbox{\columnwidth}{
        {\textbf{Takeaway:}} #1}
    }
}

\newcommand{\grammarRuleDefColor}{YellowOrange}
\newcommand{\grammarTerminalColor}{LimeGreen}
\newcommand{\grammarSymbolColor}{RawSienna}
\newcommand{\grammarDefinedNonterminalColor}{Turquoise}
\newcommand{\grammarTypeColor}{Orchid}

\author{Oğuzcan Kırmemiş}\authornote{Both authors contributed equally to the paper}
\email{oguzcan.kirmemis@gmail.com}
\orcid{0009-0003-9478-9718}
\affiliation{%
    \institution{Technical University of Munich}
    \city{Munich}
    \country{Germany}
}

\author{Francisco Romão}\authornotemark[1]

\email{francisco.romao@tum.de}
\orcid{0009-0004-0145-9785}
\affiliation{%
    \institution{Technical University of Munich}
    \city{Munich}
    \country{Germany}
}

\author{Emmanouil Giortamis}
\email{emmanouil.giortamis@in.tum.de}
\orcid{0009-0000-3638-2969}
\affiliation{%
    \institution{Technical University of Munich}
    \city{Munich}
    \country{Germany}
}

\sloppy
\author{Pramod Bhatotia}

\email{pramod.bhatotia@tum.de}
\orcid{0000-0002-3220-5735}
\affiliation{%
    \institution{Technical University of Munich}
    \city{Munich}
    \country{Germany}
}

\begin{abstract}
While the prominent quantum computing architectures are based on superconducting technology, new quantum hardware technologies are emerging, such as Trapped Ions, Neutral Atoms (or FPQAs), Silicon Spin Qubits, etc. This diverse set of technologies presents fundamental trade-offs in terms of scalability, performance, manufacturing, and operating expenses. To manage these diverse quantum technologies, there is a growing need for a {\em retargetable compiler} that can efficiently adapt existing code to these emerging hardware platforms. Such a retargetable compiler must be {\em extensible} to support new and rapidly evolving technologies, {\em performant} with fast compilation times and high-fidelity execution, and verifiable through rigorous equivalence checking to ensure the functional {\em equivalence} of the retargeted code.

To this end, we present \projectname, the first extensible, performant, and verifiable retargetable quantum compiler framework with a focus on FPQAs due to their unique, promising features. \projectname introduces \wqasm, the first formal extension of the standard OpenQASM quantum assembly with FPQA-specific instructions to support their distinct capabilities. Next, \projectname implements the \woptimizer, an extensible set of FPQA-specific optimization passes to improve execution quality. Last, the \wchecker automatically checks for equivalence between the original and the retargeted code. Our evaluation shows that \projectname improves compilation times by $10^3\times$, execution times by $4.4\times$, and execution fidelity by $10\%$, on average, compared to superconducting and state-of-the-art (non-retargetable) FPQA compilers.
\end{abstract}



\maketitle

\section{Introduction}
\label{section:introduction}

Quantum computing offers the potential to solve computational problems intractable by classical computers by leveraging the principles of quantum mechanics \cite{arute2019quantum, farhi2014quantum}, with applications in cryptography \cite{shor1999polynomial}, optimization \cite{farhi2014quantum}, chemistry \cite{kandala2017hardware, peruzzo2014variational}, machine learning \cite{biamonte2017quantum}, among others.

Quantum computing is already practically available as quantum processors are mainly offered on the cloud \cite{ibmQuantum, google-quantum, aws-quantum}. These processors are predominantly manufactured based on the Superconducting  technology~\cite{arute2019quantum}, although quantum hardware development is progressing rapidly, with various candidate technologies emerging, including Trapped Ions \cite{cirac1995quantum}, Neutral Atoms \cite{henriet2020quantum}, and Photonics \cite{o2009photonic}. 

These different quantum hardware technologies present trade-offs between performance metrics, manufacturing complexity, and operational requirements. Typical performance metrics include gate speeds, akin to clock speed in CPUs, coherence times, i.e., the time the system maintains quantum mechanical properties, and gate error rates, i.e., the probabilities of erroneous gate output \cite{google-nisq-properties}. In terms of operational requirements, some technologies require close to absolute zero temperatures \cite{krinner2019engineering, valenzuela2006microwave}, which is extremely costly, while others can operate at room-level temperatures \cite{eschner2003laser, adams1997laser}.  

Given such tradeoffs, it is evident that these diverse quantum technologies will co-exist in parallel. This coexistence creates a need for retargetable quantum compilers that can seamlessly adapt quantum programs to different hardware platforms. Specifically, retargetable compilers allow researchers and developers to leverage the unique advantages of each technology without rewriting or redesigning their quantum algorithms from scratch, thereby enhancing flexibility, accelerating development, and maximizing the utility of existing quantum resources across multiple hardware platforms.

More  specifically, a retargetable compiler must offer three fundamental properties.
{\bf First}, \textit{extensiblilty} to support new technologies and instructions in evolving technologies. {\bf Second}, \textit{performance and fidelity}, i.e., incur low compilation runtimes and optimize the code to leverage the underlying hardware's unique capabilities for improved execution fidelity. {\bf Third}, \textit{verifiability}, i.e., the compiler is rigorously checked to ensure functional equivalence to the original program.

However, designing such a retargetable compiler framework poses its own challenges. First, extensibility is non-trivial in the largely heterogeneous and evolving quantum landscape since new hardware features are being developed constantly. Second, performance is hard to achieve: fast compilation is challenging due to NP-hard compilation stages \cite{cowtan2019qubit}, and high execution fidelity is challenging due to hardware noise \cite{preskill2018quantum}. Last, checking the original and retargeted binary for functional equivalence is especially hard in quantum due to inherent randomness and computational complexity \cite{burgholzer2020advanced}.  

To target these challenges, we pose the following research question: {\em How to design an extensible, performant, and verifiable retargetable compiler framework by building on existing open standards?}

To answer this question, we propose \projectname{}, the first retargetable quantum compiler framework that automatically generates, optimizes, and verifies high-fidelity code for diverse quantum technologies. \projectname primarily focuses on two prominent technologies: superconducting qubits and neutral atoms. The former is widely studied in research and industry settings, offering a mature and comprehensive ecosystem. Neutral atoms, and specifically, Field-Programmable-Quantum-Arrays (FPQAs), show better scalability, more flexible arrangement, lower cooling requirements, and higher parallelism \cite{henriet2020quantum, saffman2010quantum, levine2019parallel, levine2018high}, compared to superconducting.

\begin{figure}
    \centering
    \includegraphics[width=\columnwidth]{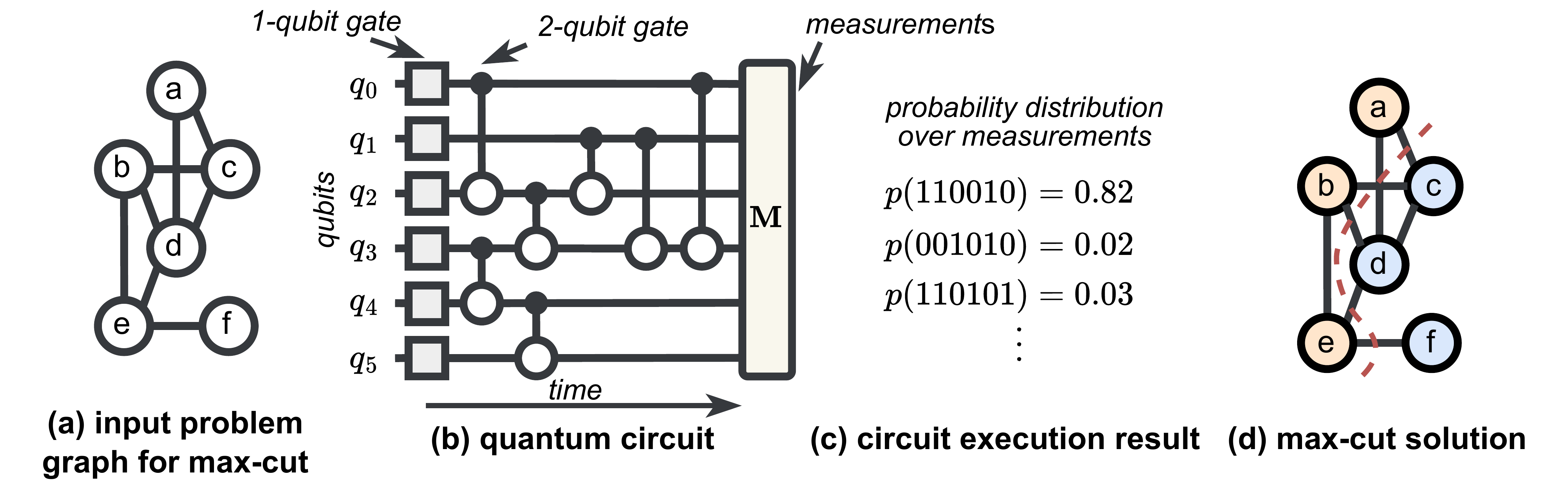}
    \caption{Example of a typical quantum algorithm (\S~\ref{sec:background:101}) {{\bf (a)} Input graph for max-cut. {\bf (b)} The quantum circuit encoding the formulation of max-cut for the graph. {\bf (c)} The result of circuit execution is a probability distribution of bitstrings. {\bf (d)} The result of (c) is interpreted as a max-cut between vertices $\{a,b,e\}$ and $\{c,d,f\}$. }}
    \label{fig:qaoa-background}
\end{figure}

We design \projectname with minimal assumptions about the hardware specifications since FPQAs are still an emerging technology that will evolve over time. Specifically, \circled{1} \projectname{} introduces the \textbf{\textit{\wqasm{}} assembly language}, the first formal extension of the standardized and widely used OpenQASM quantum assembly \cite{cross2022openqasm} with FPQA-specific backend instructions (\S~\ref{section:wqasm}). \circled{2} \projectname{} optimizes programs for FPQAs by implementing \textbf{\textit{\woptimizer{}}}, an extensible set of optimization passes that leverage FPQA capabilities to increase parallelization, reduce the program execution time, and lower the hardware noise overheads (\S~\ref{section:woptimizer}). Lastly \circled{3}, \projectname{} introduces \textbf{\textit{\wchecker{}}} which checks for equivalence between the technology-agnostic and the target-compiled binaries (\S~\ref{section:wchecker}).

We implement \projectname{} in Python on top of Qiskit \cite{Qiskit} and OpenQASM \cite{openqasm3} and PySAT \cite{imms-sat18}. We evaluate \projectname{} using state-of-the-art quantum applications on IBM quantum devices and FPQA simulators. Our results show that \woptimizer{} achieves $5.7*10^3 \times$ faster compilation times, improves execution times by $4.4\times$, and execution fidelity by $10\%$ for small applications and $1.27*10^5$ for applications of increasing size, on average, compared to superconducting architectures and state-of-the-art (non-retargetable) FPQA-specific compilers, namely \textit{Geyser} \cite{patel2022geyser}, \textit{Atomique} \cite{wang_atomique_2024}, and \textit{DPQA} \cite{tan_compiling_2024}.
\section{Background}
\label{section:background}

\subsection{Quantum Computing 101: An Example}
\label{sec:background:101}

To understand the basics of quantum computing, consider the classic max-cut problem \cite{karp2010reducibility}, which can be solved using the Quantum Approximate Optimization Algorithm (QAOA) \cite{farhi2014quantum}. QAOA is a hybrid quantum-classical algorithm that uses a quantum computer to run a parameterized quantum circuit while a classical computer optimizes the parameters. This process aims to find the optimal solution by minimizing a cost function encoded in the quantum circuit. The selection of the QAOA algorithm as a case study for \projectname is grounded in its capacity to exploit the distinctive characteristics of Max-3SAT formulations associated with NP-hard problems. These formulations are effectively represented within a QAOA circuit designed for execution on FPQA architectures.

\begin{figure*} [ht]
    \centering
    \includegraphics[width=0.75\textwidth]{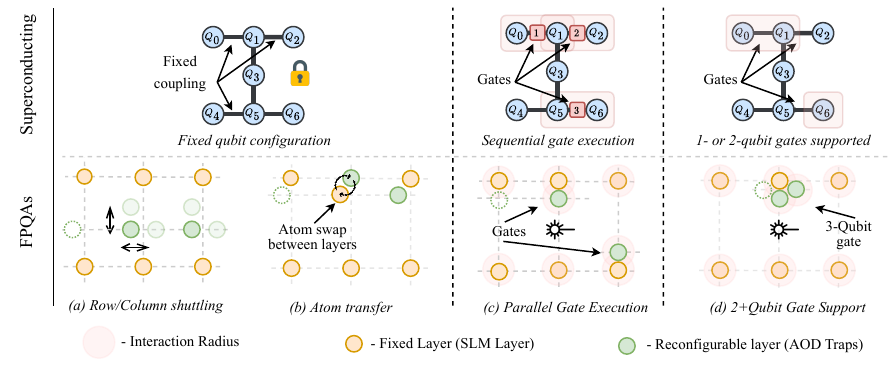}
    \caption{Unique capabilities of FPQA hardware (\S~\ref{sec:quantum_architectures}). {FPQAs allow shuttling rows and/or columns of the reconfigurable atom layer, transferring atoms, executing gates in parallel, and native multi-qubit gates support.}}
    \vspace{-2mm}
    \label{fig:fpqa_capabilities}
\end{figure*}

Figure \ref{fig:qaoa-background} shows how QAOA solves a max-cut problem on an example graph (a). The problem is encoded as a quantum circuit (b), with each qubit representing a vertex of the input graph. Quantum gates are applied over time to change the state of the qubits, and at the end, measurements provide bitstrings as output. 
Unlike classical circuits, quantum circuits are probabilistic due to the superposition property of qubits. Quantum gates have probabilistic effects, and the final result is obtained by executing the circuit multiple times. The solution is a probability distribution over all possible bitstrings of the measured qubits, as shown in Figure~\ref{fig:qaoa-background} (c).

In our example, the probability distribution represents the max-cut problem's solution, with high probabilities indicating potential solutions. In Figure \ref{fig:qaoa-background} (d), the solution is the bitstring \texttt{110010}, which has the highest probability. This means qubits $q_0$, $q_1$, and $q_4$ are measured as 1, placing vertices $\{a,b,e\}$ in one partition and vertices $\{c,d,f\}$ in the other.

\subsection{QPUs and Performance Metrics}
\label{section:background:technical-foundations}

\myparagraph{QPU characteristics}
Today's Quantum Processing Units (QPUs) are categorized as noisy intermediate-scale quantum (NISQ) devices \cite{preskill2018quantum} due to their limited number of qubits (up to a few hundred \cite{ibmQuantum}) and their vulnerability to hardware and environmental noise \cite{georgopoulos2021modeling}. Specifically, qubit measurements can result in bit-flip errors, and gate operations may perform incorrectly \cite{google-nisq-properties}. Additionally, qubits tend to collapse from the quantum (superposition) state and behave like classical bits (\textit{decoherence effect} \cite{klimov2018fluctiations}) and destructively interfere with each other through \textit{crosstalk effects} \cite{cross2019validating}. Last, QPU qubits can interact with each other only if they are connected with a physical link, with some QPUs having static connectivity layouts, while others have reconfigurable \cite{arute2019quantum, henriet2020quantum}.

\myparagraph{Fidelity performance metric}
Fidelity \cite{fidelity-qiskit} is normally used to assess a circuit's performance on noisy QPUs, ranging from $[0,1]$, comparing the noisy and ideal outputs. Similarly, we use the Estimated Probability of Success (EPS), which is the probability of one run outputting the correct result.

\subsection{Quantum Hardware Architectures}
\label{sec:quantum_architectures}

\myparagraph{Quantum hardware diversity}
Currently, several candidate technologies are being actively developed, including superconducting qubits \cite{arute2019quantum}, trapped ions \cite{cirac1995quantum}, photonics \cite{o2009photonic}, and Field-Programmable Quantum Arrays (FPQAs) \cite{henriet2020quantum}. Each technology's characteristics differ (\S~\ref{section:background:technical-foundations}), rendering it more suitable for certain applications than others.

\myparagraph{Prominent technologies}
\projectname targets two quantum technologies: superconducting qubits and FPQAs. Superconducting technology has the advantages of fast gate execution \cite{siddiqi2021engineering, dicarlo2009demonstration, majer2007coupling}, but mostly the ease of integration, maturity, and software support, compared to other architectures. However, it also suffers from low and rigid qubit connectivity, low coherence times, and high gate errors \cite{tannu2019mitigating, murali2020software, prakash2019noise}. This is shown in Figure \ref{fig:fpqa_capabilities} (top), where the example qubit configuration is fixed.
In contrast, FPQA hardware is gaining traction due to its longer coherence times, higher resistance to environmental noise, and flexible qubit connectivity \cite{henriet2020quantum, saffman2010quantum, levine2019parallel, levine2018high}.

\myparagraph{FPQA-unique capabilities}
\label{sec:fpqa_capabilities}

Figure \ref{fig:fpqa_capabilities} presents four unique capabilities of FPQAs that show their advantage over superconducting qubits. First, FPQAs support dynamic qubit configuration and connectivity, referred to as qubit shuttling (Figure \ref{fig:fpqa_capabilities} (a)). FPQAs provide a fixed layer (yellow atoms) of traps and a moving layer (green atoms) that allows row or column shuttling, thus reconfiguring the qubit connectivity graph. Second, FPQAs enable the swapping of two atoms between two layers, i.e., atom transfer (Figure \ref{fig:fpqa_capabilities} (b)). This allows reconfiguring the connectivity graph without adding additional overheads on the quantum circuit, as done on superconducting technology. Third, FPQAs implement multi-qubit gates with global pulses (Figure \ref{fig:fpqa_capabilities} (c)), which allows parallel multi-qubit gate execution. Lastly,  
FPQAs natively support higher-order gates (Figure \ref{fig:fpqa_capabilities} (d)). Specifically, multi-qubit gates are not limited to 2-qubit interactions, being able to entangle three or more qubits.

\myparagraph{FPQA opportunities and challenges}
These features come with some challenges: first, movable rows or columns cannot move over one another, which means that for complete movement of qubits, we need to swap qubits back and forth between the fixed and moving layers; secondly, the global pulses apply gates in qubits that are close to each other, it thus, however, means that if two qubits are not supposed to interact, they need to be separated. Thirdly, gate execution is relatively slow; to counter this problem, the compiler should aim for high parallelization of gate execution.

\begin{figure*}[ht]
    \centering
    \includegraphics[width=0.9\textwidth]{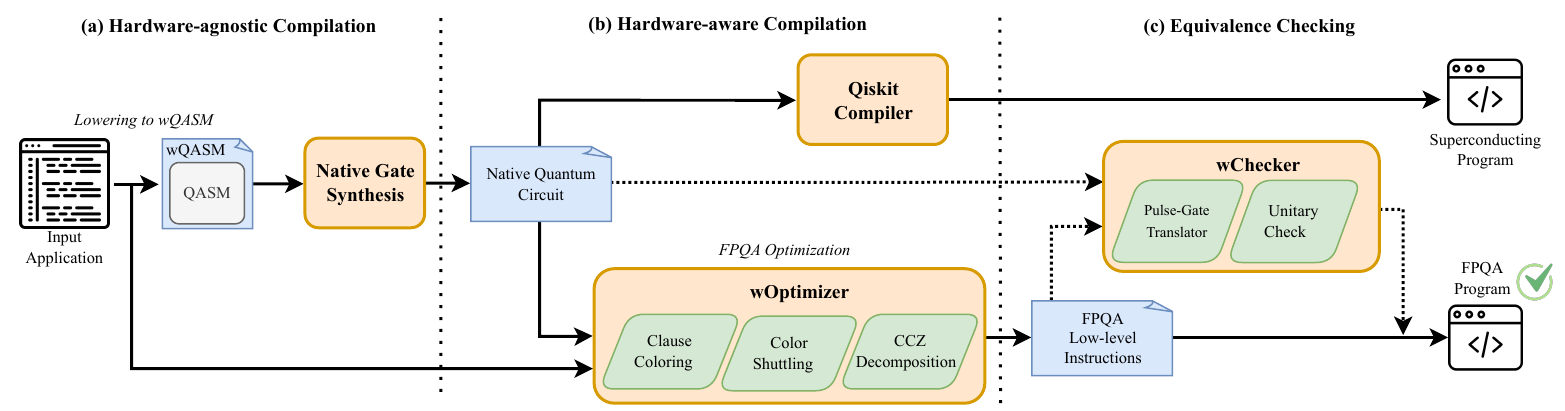}
    \caption{\projectname overview (\S~\ref{section:overview}). {\projectname's workflow consists of three stages: (a) hardware-agnostic compilation, (b) hardware-aware compilation, and (c) equivalence checking. The hardware-agnostic compilation lowers the input application to \wqasm. In the hardware-aware compilation stage, the compilation passes are specific to the target technology chosen by the user. Finally, equivalence checking ensures that the hardware-aware optimizations maintain the equivalence to the initial program.}}
    \label{fig:weaver-overview}
    \vspace{-1mm}
\end{figure*}

\section{\projectname Overview}
\label{section:overview}

Quantum technologies are constantly emerging and evolving, and specific technologies such as FPQAs are gaining traction because of their opportunities for performance improvement compared to the widely adopted superconducting architectures. However, designing a retargetable compiler that supports diverse quantum technologies while leveraging their strengths presents its own challenges.

\subsection{Design Challenges}
\label{section:overview:design_challenges}

\myparagraph{Challenge \#1: Extensibility}
A retargetable compiler must be extensible by supporting new quantum technologies, optimization passes, and instruction sets within a specific technology. To achieve this, it is crucial to adopt a standard and widely used common abstraction that acts as a common denominator across the compiler and other existing infrastructure.

\myparagraph{Challenge \#2: Performance}
A retargetable compiler must be performant by leveraging heuristics for the NP-hard compilation steps to reduce compilation time and leveraging the hardware features to improve fidelity. This requires deploying efficient and target-specific optimizations that exploit the unique characteristics of each quantum device.

\myparagraph{Challenge \#3: Equivalence checking}
A retargetable compiler must automatically check the functional equivalence of the retargeted circuit. The challenge lies in 
the \textit{exponential} memory and computation requirements to represent and simulate quantum states using classical computers. Moreover, the inherent heterogeneity of quantum technologies introduces additional nondeterminism in the program outputs.

\subsection{\projectname Retargetable Compiler Framework}
Based on the previous key ideas, we present the design of our retargetable \projectname compiler framework. At a high level, \projectname extends the OpenQASM language and uses it as an IR to apply general-purpose optimizations. Then, \projectname leverages the FPQA-unique functionalities by implementing an extensible set of optimization passes that improve circuit duration, the number of operations, and overall fidelity. Lastly, \projectname  verifies the correctness of the optimized circuit against the original one. The architecture of our compiler comprises three stages, as shown in Figure \ref{fig:weaver-overview}.

\myparagraph{Hardware-agnostic compilation} Initially, a user submits an application to \projectname as a quantum circuit. \projectname lowers the circuit to \wqasm IR and generates a \textit{native} circuit using a gate set compatible with superconducting and FPQA technologies. We detail the \wqasm extended instructions, the \wqasm grammar, and the \wqasm semantics in \S~\ref{section:wqasm}.

\myparagraph{Hardware-aware compilation} Depending on the user's backend choice, \projectname directs the native circuit to either the superconducting or the FPQA path. The superconducting (top arrow) path submits the circuit through the Qiskit Compiler \cite{qiskit-transpiler}, which fully compiles it to the superconducting backend. In the FPQA path (bottom-arrow), \projectname optimizes the circuit through the \wcompiler, which applies three sequential optimization passes we detail in \S~\ref{section:woptimizer}. Finally, the circuit is converted to FPQA pulse instructions and is ready to be submitted to FPQA hardware controllers.

\myparagraph{Equivalence checking} Lastly, during equivalence checking, \projectname submits the list of low-level FPQA pulses to the \wchecker to ensure that the optimization passes maintain the behavior of the original nativized circuit. The \wchecker comprises two steps, namely pulse-gate translator and unitary check, which we detail in \S~\ref{section:wchecker}.
\section{\wqasm Extensions}
\label{section:wqasm}

\wqasm extends OpenQASM for FPQAs with annotations that supply additional information on OpenQASM statements, defining FPQA-specific steps required before each statement.

\myparagraph{OpenQASM as the foundation}
OpenQASM (Open Quantum Assembly Language) \cite{openqasm3} is designed with an assembly-like syntax targeting quantum algorithms and hardware operations. It is extensible through pragmas and annotations, which are compiler directives providing extra information for optimization or hardware-specific requirements. Pragmas are order-independent, while annotations specifically relate to the following OpenQASM statement. We opt for OpenQASM as our IR over other IRs for two reasons: (1) \textbf{Broad adoption and industry support}: OpenQASM is widely adopted in the quantum computing community and supported by major quantum platforms. A good example is the large number of benchmark circuits written in OpenQASM \cite{quetschlich2022mqt, tomesh2022supermarq, li2021qasmbench}. (2) \textbf{Easily extensible}: OpenQASM supports pragmas and annotations, allowing extending the standard instructions.

\subsection{\wqasm FPQA Instructions}

OpenQASM is designed to be hardware-agnostic. FPQA technology provides hardware flexibility and features that instructions can control. This motivates the creation of an annotation-based extension that does not change its hardware-agnostic characteristic. Besides the functionalities presented in \S~\ref{fig:fpqa_capabilities}, FPQAs also provide two control pulses: \textbf{Raman pulse} is an atom-directed pulse that applies single-atom rotations around the $x$, $y$ or $z$ axis. This pulse can be applied to one atom or globally by sending it to every initialized atom trap. \textbf{Rydberg pulse} is a global pulse that applies a multi-atom operator to all atoms close enough to each other within the Rydberg distance. The operation applied is a controlled-Z or multi-controlled-Z gate, depending on the number of interacting atoms.

\subsection{\wqasm Grammar}

\begin{figure}
    \fontsize{8}{9}\selectfont
    \setlength{\grammarparsep}{6pt plus 1pt minus 1pt}
    \begin{grammar}
    
    <\textbf{\textcolor{\grammarRuleDefColor}{program}}> ::= <version> \textcolor{\grammarSymbolColor}{?} <statementOrScope>\textcolor{\grammarSymbolColor}{*}
    
    <\textbf{\textcolor{\grammarRuleDefColor}{version}}> ::= \textcolor{\grammarTerminalColor}{‘"OpenQASM"’} <versionSpecifier> \textcolor{\grammarTerminalColor}{‘";"’}
    
    <\textbf{\textcolor{\grammarRuleDefColor}{statementOrScope}}> ::= <statement> \, \textcolor{\grammarSymbolColor}{|} \, <scope>
    
    <\textbf{\textcolor{\grammarRuleDefColor}{scope}}> ::= \textcolor{\grammarTerminalColor}{‘"\{"’} <statementOrScope> \textcolor{\grammarSymbolColor}{*} \textcolor{\grammarTerminalColor}{‘"\}"’}
    
    <\textbf{\textcolor{\grammarRuleDefColor}{statement}}> ::= <pragma>
        \textcolor{\grammarSymbolColor}{\alt} <annotation>\textcolor{\grammarSymbolColor}{*} \textcolor{\grammarSymbolColor}{(}
            \\ \textcolor{\grammarSymbolColor}{\textbar} \quad <ioDeclarationStatement>
            \\ \textcolor{\grammarSymbolColor}{\textbar} \quad <gateStatement>
            \\ \textcolor{\grammarSymbolColor}{\textbar} \quad <gateCallStatement>
            \\ \textcolor{\grammarSymbolColor}{\textbar} \quad "..."
        \\ \llap{\textcolor{\grammarSymbolColor}{)}\quad}
    
    <\textbf{\textcolor{\grammarRuleDefColor}{annotation}}> ::= <slmDefinition>
        \textcolor{\grammarSymbolColor}{\alt} <aodDefinition>
        \textcolor{\grammarSymbolColor}{\alt} <atomBind>
        \textcolor{\grammarSymbolColor}{\alt} <trapTransfer>
        \textcolor{\grammarSymbolColor}{\alt} <aodShuttle>
        \textcolor{\grammarSymbolColor}{\alt} <raman>
        \textcolor{\grammarSymbolColor}{\alt} <rydberg>
        \textcolor{\grammarSymbolColor}{\alt} <annotationKeyword> <remainingLineContent>\textcolor{\grammarSymbolColor}{?}
    
    <\textbf{\textcolor{\grammarRuleDefColor}{slmDefinition}}> ::= \textcolor{\grammarTerminalColor}{‘"@slm"’} <trapPositions>

    <\textbf{\textcolor{\grammarRuleDefColor}{trapPositions}}> ::= \textcolor{\grammarTerminalColor}{‘"["’} <position> \textcolor{\grammarSymbolColor}{(} \textcolor{\grammarTerminalColor}{‘","’} <position> \textcolor{\grammarSymbolColor}{)}\textcolor{\grammarSymbolColor}{*} \textcolor{\grammarTerminalColor}{‘"]"’}

    <\textbf{\textcolor{\grammarRuleDefColor}{position}}> ::= \textcolor{\grammarTerminalColor}{‘"("’} \syntleft\textcolor{\grammarTypeColor}{float}\syntright  \textcolor{\grammarTerminalColor}{‘","’} \syntleft\textcolor{\grammarTypeColor}{float}\syntright  \textcolor{\grammarTerminalColor}{‘")"’} 
    
    <\textbf{\textcolor{\grammarRuleDefColor}{aodDefinition}}> ::= \textcolor{\grammarTerminalColor}{‘"@aod"’} <aodRows> <aodColumns>
    
    <\textbf{\textcolor{\grammarRuleDefColor}{atomBind}}> ::= \textcolor{\grammarTerminalColor}{‘"@bind"’} \syntleft\textcolor{\grammarTypeColor}{identifier}\syntright 
        \\  \textcolor{\grammarSymbolColor}{(} \textcolor{\grammarTerminalColor}{‘"slm"’} \syntleft\textcolor{\grammarTypeColor}{integer}\syntright  \, \textcolor{\grammarSymbolColor}{|} \, \textcolor{\grammarTerminalColor}{‘"aod"’} \syntleft\textcolor{\grammarTypeColor}{integer}\syntright  \syntleft\textcolor{\grammarTypeColor}{integer}\syntright \textcolor{\grammarSymbolColor}{)} 
    
    <\textbf{\textcolor{\grammarRuleDefColor}{atomTransfer}}> ::= \textcolor{\grammarTerminalColor}{‘"@transfer"’} \syntleft\textcolor{\grammarTypeColor}{integer}\syntright  
        \\ \textcolor{\grammarTerminalColor}{‘"("’} \syntleft\textcolor{\grammarTypeColor}{integer}\syntright  \textcolor{\grammarTerminalColor}{‘","’} \syntleft\textcolor{\grammarTypeColor}{integer}\syntright  \textcolor{\grammarTerminalColor}{‘")"’}
    
    <\textbf{\textcolor{\grammarRuleDefColor}{aodShuttle}}> ::= \textcolor{\grammarTerminalColor}{‘"@shuttle"’} 
        \\ \textcolor{\grammarSymbolColor}{(} \textcolor{\grammarTerminalColor}{‘"row"’} \, \textcolor{\grammarSymbolColor}{|} \, \textcolor{\grammarTerminalColor}{‘"column"’} \textcolor{\grammarSymbolColor}{)} \syntleft\textcolor{\grammarTypeColor}{integer}\syntright  \syntleft\textcolor{\grammarTypeColor}{integer}\syntright 
    
    <\textbf{\textcolor{\grammarRuleDefColor}{raman}}> ::= \textcolor{\grammarTerminalColor}{‘"@raman global"’} \syntleft\textcolor{\grammarTypeColor}{float}\syntright  \syntleft\textcolor{\grammarTypeColor}{float}\syntright  \syntleft\textcolor{\grammarTypeColor}{float}\syntright 
        \textcolor{\grammarSymbolColor}{\alt} \textcolor{\grammarTerminalColor}{‘"@raman local"’} \syntleft\textcolor{\grammarTypeColor}{identifier}\syntright  \syntleft\textcolor{\grammarTypeColor}{float}\syntright \syntleft\textcolor{\grammarTypeColor}{float}\syntright  \syntleft\textcolor{\grammarTypeColor}{float}\syntright 
    
    <\textbf{\textcolor{\grammarRuleDefColor}{rydberg}}> ::= \textcolor{\grammarTerminalColor}{‘"@rydberg"’}

    <"\textcolor{\grammarRuleDefColor}{...}"> ::= "..."
    
    \end{grammar}
    \hfill
    \hfill
    \caption{Abstract grammar for \wqasm in EBNF format. Note that the non-terminals highlighted in purple are renamed from the OpenQASM grammar for simplification purposes. Their definitions, the remaining rules, and the full version of the OpenQASM grammar can be found in OpenQASM specifications~\cite{openqasm2,openqasm3,openqasm3grammar}.}
    \label{fig:wqasm-grammar}
\end{figure}

\wqasm extends OpenQASM grammar rules with FPQA instructions as a superset of OpenQASM, detailed in Figure~\ref{fig:wqasm-grammar}. In \wqasm, OpenQASM instructions are annotated with their FPQA equivalents for direct FPQA programming. Challenges arise when FPQA instructions like shuttling don't directly correspond to logical gates and depend on the FPQA's previous atoms' positions. For instance, Rydberg pulses apply a transformation to atoms that were moved together by previous shuttling instructions. These non-pulse FPQA actions are then associated with the respective logical gates. A global Raman pulse also translates into a logical gate affecting all qubits, while a local Raman pulse uses a single $U3$ gate.

The \wqasm file includes redundancy between the logical gate instructions from the original OpenQASM file and the FPQA-specific annotations. The annotations specify steps for each logical gate that must be executed sequentially, as each step depends on the previous state of the FPQA device. In contrast, logical gate instructions can be executed in parallel if their dependencies are met and they do not share qubits, following the order dictated by a dependency graph. This flexibility does not apply to FPQA annotations requiring a fixed sequence. During compilation, \projectname determines the execution order of the gates, and the \wchecker then checks that the FPQA annotations correctly implement the same logical circuit as intended by the original OpenQASM code. Once the challenges are addressed, \wqasm files can be treated like regular OpenQASM files (ignoring FPQA annotations). This allows them to be retargeted to other quantum architectures, possibly with additional compilation for specific hardware.

\subsection{\wqasm Semantics} 
We formalize the \wqasm semantics and thoroughly explain its objective, arguments, pre-conditions, and the outcome of that instruction (post-condition). The proposed annotations are summarized in Table \ref{tab:openqasm}. 

\begin{table*}[t]
\centering
\caption{Annotations list for extending OpenQASM to FPQA technology.}
\fontsize{8}{9}\selectfont
\begin{tabular}{p{0.08\textwidth}p{0.14\textwidth}p{0.18\textwidth}p{0.28\textwidth}p{0.21\textwidth}}
\toprule
\textbf{Annotation} & \textbf{Arguments} & \textbf{Description} & \textbf{Pre-condition} & \textbf{Post-condition}  \tabularnewline
\hline
\multirow{1}{*}{\texttt{@slm}} & \texttt{[(x$_0$,y$_0$),}...\texttt{,(x$_n$,y$_n$)]} & Distribute SLM atom trap locations on input coordinates & \texttt{Dist}((\texttt{x$_i$}, \texttt{y$_i$}), (\texttt{x$_j$}, \texttt{y$_j$})) > \texttt{Dist}$_\texttt{min}$, \texttt{i $\neq$ j} & (\texttt{SLMX$_i$}, \texttt{SLMY$_i$}) = (\texttt{x$_i$}, \texttt{y$_i$}) \tabularnewline
\midrule
\multirow{1}{*}{\texttt{@aod}} & \texttt{[x$_0$,} ...\texttt{x$_n$]} \newline \texttt{[y$_0$,} ...\texttt{, y$_n$]} & Setup AOD atom-trap grid based on input coordinates & \texttt{Dist}((\texttt{x$_i$}, \texttt{y$_i$}), (\texttt{x$_j$}, \texttt{y$_j$})) > \texttt{Dist$_\texttt{min}$} \newline \texttt{x$_\texttt{n}$} < \texttt{x$_\texttt{(n+1)}$} & (\texttt{AODX$_i$}, \texttt{AODYW$_i$}) = (\texttt{x$_i$}, \texttt{y$_i$}) \tabularnewline
\midrule
\multirow{1}{*}{\texttt{@bind}} & \texttt{q$_\texttt{id}$ \newline slm/aod \newline slm$_\texttt{index}$/(aod$_\texttt{x}$,aod$_\texttt{y}$)} & Bind qubit-indexes in SLM with \texttt{index} or AOD atom-traps at \texttt{x} and \texttt{y} & None & (\texttt{AODX$_\texttt{index}$}, \texttt{AODY$_\texttt{index}$)} $\leftarrow$ \texttt{q$_\texttt{id}$} or (\texttt{SLMX$_i$}, \texttt{SLMY$_i$}) $\leftarrow$ \texttt{q$_\texttt{id}$} \tabularnewline
\midrule
\multirow{1}{*}{\texttt{@transfer}} & \texttt{slm$_\texttt{index}$} \newline (\texttt{aod$_\texttt{x}$}, \texttt{aod$_\texttt{y}$}) & Transfer atom from SLM trap at \texttt{index} to AOD at \texttt{x} and \texttt{y} & \texttt{Dist}(\texttt{SLM$_\texttt{index}$}, (\texttt{AODX}, \texttt{AODY})) < \texttt{Dist\_Transfer$_\texttt{Max}$} & (\texttt{AODX}, \texttt{AODY})' $\dashleftarrow$ \texttt{SLM$_\texttt{index}$} \newline  \texttt{SLM$_\texttt{index}$}' $\dashleftarrow$ (\texttt{AODX}, \texttt{AODY}) \tabularnewline
\midrule
\multirow{1}{*}{\texttt{@shuttle}} & \texttt{row/column \newline index \newline offset} &
Move AOD row or column at \texttt{index} by \texttt{offset} & \texttt{Dist}(\texttt{AODX$_\texttt{i}$}, \texttt{AODX$_\texttt{i+1}$} OR \texttt{AODX$_\texttt{i-1}$}) > \texttt{Dist$_\texttt{min}$} \newline \texttt{Dist}(\texttt{AODY$_\texttt{i}$}, \texttt{AODY$_\texttt{i+1}$} OR \texttt{AODY$_\texttt{i-1}$}) > \texttt{Dist$_\texttt{min}$} & \texttt{AODX$_\texttt{index}$} + \texttt{offset} OR \texttt{AODY$_\texttt{index}$} + \texttt{offset} \tabularnewline
\midrule
\multirow{1}{*}{\texttt{@raman local}} & \texttt{q$_\texttt{id}$ \newline $(x,y,z)$} & Rotate \texttt{q$_\texttt{id}$} with angles \texttt{x}, \texttt{y} and \texttt{z} & \texttt{q$_\texttt{id}$} $\neq$ \texttt{None} & \texttt{R}(\texttt{x}, \texttt{y}, \texttt{z}) \ket{\texttt{q$_\texttt{id}$}} \tabularnewline
\multirow{1}{*}{\texttt{@raman global}} & $(x,y,z)$ & Rotate all qubits with angles \texttt{x}, \texttt{y} and \texttt{z} & None & R(\texttt{x}, \texttt{y}, \texttt{z}) \ket{\texttt{q$_\texttt{id}$}}\ \ \ $\forall\ \texttt{id}$ \tabularnewline
\midrule
\multirow{1}{*}{\texttt{@rydberg}} & None & Apply CZ on qubits closer than the Rydberg distance & None  & If \texttt{Dist}(\texttt{q}$_\texttt{i}$, \texttt{q}$_\texttt{j}$) $\leq$ \texttt{Rydberg\_dist}: \newline \texttt{CZ} \ket{\texttt{q$_\texttt{i}$}\ \texttt{q$_\texttt{j}$}}\ \ \ $\forall$ \texttt{i,j; $\texttt{i} \neq \texttt{j}$} \tabularnewline
\bottomrule
\end{tabular}

\vspace{-2mm}
\label{tab:openqasm}
\end{table*}

\myparagraph{Initialization of SLM Traps (\texttt{@slm})} Configures a fixed layer of atom traps at specified coordinates. Each coordinate \texttt{(x$_i$, y$_i$)} represents the position of an atom trap in a 2D plane.
\begin{itemize}[leftmargin=*]
    \item {\em Pre-condition:} Input coordinates need to be at a minimum distance from each other to avoid unwanted qubit inference. This distance is usually between 5 to 10 micrometers.
    \item  {\em Post-condition:} Fixed layer of atom traps is initialized with traps on locations \texttt{(x$_i$, y$_i$)}.
\end{itemize}

\myparagraph{Initialization of AOD Traps (\texttt{@aod})} Set up a reconfigurable-layer of atom traps. It takes the coordinates arrays \texttt{x} and \texttt{y}, allowing to move trap locations. In contrast to the fixed layer layout, where atoms can be placed in arbitrary locations, the AOD layer is always in a grid-like format.
\begin{itemize}[leftmargin=*]
    \item {\em Pre-condition:} Values for the x and y coordinates must be in increasing order, and a minimum distance must be preserved between adjacent rows and columns.
    \item {\em Post-condition:} AOD layer is initialized with rows and columns on coordinates given by \texttt{x} and \texttt{y}, respectively.
\end{itemize}

\myparagraph{Binding atoms to qubit IDs (\texttt{@bind})} Physical traps are bound to unique qubit IDs; needed for the compilation process.
\begin{itemize}[leftmargin=*]
    \item {\em Pre-condition:} No pre-conditon.
    \item {\em Post-condition:} Atom at \texttt{slm$\_\texttt{index}$} on a SLM layer, or \texttt{aod\_\texttt{x}}, \texttt{aod\_\texttt{y}} on a AOD layer is now tied to qubit \texttt{q\_\texttt{id}}.
\end{itemize}

\myparagraph{Atom transfer between layers (\texttt{@transfer})} Transfers an atom from the SLM layer to the AOD layer, or vice versa.
\begin{itemize}[leftmargin=*]
    \item {\em Pre-condition:} Destination atom trap needs to be empty for the atom to be moved onto. The involved atom traps need to be close enough.
    \item {\em Post-condition:} An atom is moved between layers.
\end{itemize}

\myparagraph{Moving the reconfigurable layer (\texttt{@shuttle})}

This instruction applies a move operation, either to a row or column.
\begin{itemize}[leftmargin=*]
    \item {\em Pre-condition:} Adjacent rows and columns need to maintain a minimum distance (5-10 micrometers). The target row/column should not move over any other row/column.
    \item {\em Post-condition:} Row or column \texttt{index} moved by \texttt{offset}.
\end{itemize}

\myparagraph{Local Raman pulse (\texttt{@raman local})} Applies a Raman local pulse on \texttt{q$_\texttt{id}$}.
\begin{itemize}[leftmargin=*]
    \item {\em Pre-condition:} Qubit with ID \texttt{q$_\texttt{id}$} exists.
    \item {\em Post-condition:} Qubit with ID \texttt{q$_\texttt{id}$} is rotated by \texttt{x}, \texttt{y} and \texttt{z} around the axis $x$, $y$ and $z$, respectively.
\end{itemize}

\myparagraph{Global Raman pulse (\texttt{@raman global})} Applies a global Raman pulse with rotations $(x,y,z)$.
\begin{itemize}[leftmargin=*]
    \item {\em Pre-condition:} No pre-condition.
    \item {\em Post-condition:} All qubits from both layers are rotated by \texttt{(x, y, z)} around the $x$, $y$ and $z$ axis.
\end{itemize}

\myparagraph{Rydberg pulse (\texttt{@rydberg})} Applies a global Rydberg pulse.
\begin{itemize}[leftmargin=*]
    \item {\em Pre-condition:} No pre-condition.
    \item {\em Post-condition:} All atoms within a Rydberg radius of each other are applied a multi-qubit controlled $Z$ gate.
\end{itemize}
\section{\woptimizer Optimizer}
\label{section:woptimizer}

The \woptimizer module focuses on optimizing FPQA device-targeted circuits, particularly QAOA circuits addressing combinatorial optimization challenges like the Travelling Salesman Problem, Maximum Cut Problem (MaxCut), or Maximum 3-Satisfiability Problem (MAX-3SAT), crucial in operations research and cost reduction in sectors such as logistics~\cite{phillipson2024quantumcomputinglogisticssupply}. 

The Quantum Approximate Optimization Algorithm is a prominent algorithm in the NISQ era, favored for its minimal connectivity needs. It's considered influential towards achieving practical quantum supremacy~\cite{isgc_2019} due to its applicability on NISQ-era hardware. QAOA circuits consist of three parts: initialization of a mixer Hamiltonian, the time evolution of a cost Hamiltonian embedding a given optimization problem, and the time evolution of the mixer Hamiltonian itself. Our optimizations target the implementation of the time evolution of the cost Hamiltonian, as introduced below.

\myparagraph{Our approach}
\woptimizer focuses on MAX-3SAT problems for the FPQA architecture. This decision is not restrictive, as all NP problems can be reduced to MAX-3SAT. Additionally, the optimization techniques we introduce here can be adapted to general QAOA circuits with some restrictions.

\begin{figure*}
    \centering
    \includegraphics[width=0.9\textwidth]{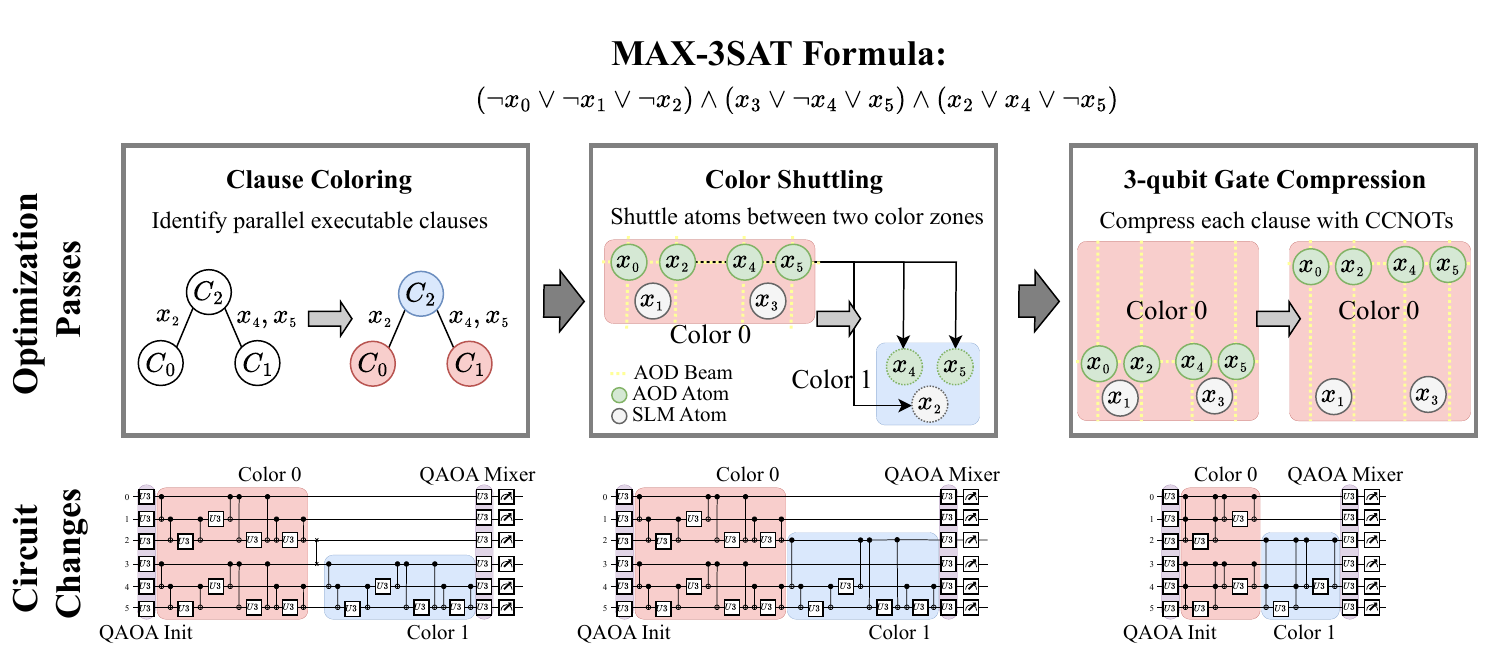}
    \caption{Overview of the optimization passes done by \woptimizer: \emph{clause coloring}, \emph{color shuttling}, \emph{3-qubit gate compression}. \emph{Clause coloring} identifies the clauses that can be executed in parallel to utilize the global lasers in FPQA. \emph{Color shuttling} eliminates the swap overhead between executing two different colors. \emph{3-qubit gate compression} shortens the subcircuit fragments for each clause from 8 $CNOT$ to 2 $CNOT$ and 2 $CCNOT$ gates. This reduces the total number of required pulses, as each $CNOT$/$CCNOT$ gate can be implemented with a $CZ$/$CCZ$ gate that the FPQA natively supports.}
    \label{fig:woptimizer-overview}
    \vspace{-2mm}
\end{figure*}

\begin{figure}
    \centering
    \begin{adjustbox}{width=0.9\linewidth}
    \begin{subfigure}{0.3\textwidth}
        \begin{quantikz}
            & \qw & \qw & \ctrl{1} & \qw & \ctrl{1} & \qw & \qw \\
            & \qw & \qw & \targ{} & \gate{R_Z(\gamma)} & \targ{} & \qw & \qw  \\
        \end{quantikz}
        \caption{}
        \label{fig:rzz_implementation}
    \end{subfigure}
    \begin{subfigure}{0.3\textwidth}
        \begin{quantikz}
            & \ctrl{1} & \qw & \qw & \qw & \ctrl{1} & \qw \\
            & \targ{} & \ctrl{1} & \qw & \ctrl{1} & \targ{} & \qw \\
            & \qw & \targ{} & \gate{R_Z(\gamma)} & \targ{} & \qw & \qw \\
        \end{quantikz}
        \caption{}
        \label{fig:rzzz_implementation}
    \end{subfigure}
    \end{adjustbox}
    \caption{Compilation of terms in the objective polynomial function of MAX-3SAT formula. \textbf{(\ref{fig:rzz_implementation})} \emph{Circuit fragment for a quadratic term}. \textbf{(\ref{fig:rzzz_implementation})} \emph{Implementation of a cubic term}. Terms with single variables are not shown in this figure, but they are simply compiled as a $R_Z$ gate with angle $\gamma$.}
    \label{fig:rzz_rzzz_implementation}
\end{figure}
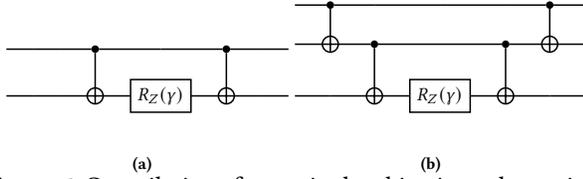

\myparagraph{Workflow using an example}
Figure~\ref{fig:woptimizer-overview} illustrates an example formula where the cost Hamiltonian aims to maximize satisfied clauses. An example clause, $(\neg x_0 \lor \neg x_1 \lor \neg x_2)$, is represented as $f(x_0, x_1, x_2) = -x_0 x_1 x_2$. The overall formula aggregates these clause-specific objective functions into a Boolean polynomial, limiting terms to cubic degrees. For quantum circuit implementation, terms convert to z-axis rotations via a $CNOT$ ladder configuration, depicted in Figure~\ref{fig:rzz_rzzz_implementation}.

\begin{algorithm}[t]
    \caption{Pseudocode for clause coloring.}
    \fontsize{8}{9}\selectfont
    \label{alg:clause_coloring}
    \begin{algorithmic}
        \Require $F \gets$ List of clauses \Comment{[[-1, -2, -3], [4, -5, 6], [3, 5, -6]]}
        \Ensure $C \gets$ Color assignments \Comment{[0, 0, 1]}
        \State $n \gets$ Number of clauses
        \State $V \gets$ List of clauses \Comment{[0, 1, 2]}
        \State $E[n][n] \gets$ Initialized to zero \Comment{Adjacency matrix}
        \For{$C_i$, $C_j$ in $F$ and $C_i \neq C_j$}
            \If{$C_i \cap C_j \neq \emptyset$}
                \State $E[i][j] = 1$
            \EndIf
        \EndFor
        \State \Return $C \gets \texttt{DSatur(V, E)}$
    \end{algorithmic}
\end{algorithm}

\subsection{\projectname Optimization Overview}
\woptimizer consists of three stages, each targeting a different goal. Figure~\ref{fig:woptimizer-overview} shows the overview of the optimizations. Each stage takes advantage of FPQA's unique hardware features to develop an optimized circuit. These stages are:

\begin{itemize}[leftmargin=*]
    \item \textbf{Clause coloring:} Identifies independent clauses to maximize the parallelization capabilities of FPQAs.
    \item \textbf{Color shuttling:} Instead of a swap gate-based routing approach, we use the shuttling mechanism to eliminate the swap overhead in the circuit execution completely.
    \item \textbf{3-qubit gate compression:} Multi-qubit gates in each clause get compressed into fragments with 3-qubit gates that are natively supported by FPQAs, reducing the overall circuit depth and gate count.
\end{itemize}

\noindent
The following sections illustrate the effects of each stage with the running example in Figure~\ref{fig:woptimizer-overview}.

\subsection{Clause Coloring}
\label{section:woptimizer:clause_coloring}

\myparagraph{Challenge} To offset FPQA's slow gate execution times, we maximize parallelization by increasing multi-qubit gate operations with Rydberg pulses. In MAX-3SAT, we notice that the cost Hamiltonian circuits of two independent clauses can be executed in parallel. Thus, we can divide the formula into independent clause clusters to decrease their total number.

\myparagraph{Key idea} Clauses' relationships are represented as an undirected graph, with edges between clauses sharing a variable. Clustering becomes a graph coloring problem, where each node is assigned a color so that two neighboring nodes always get assigned different colors. Thus, clauses with the same color can be executed in parallel. For example, in Figure~\ref{fig:woptimizer-overview}, we run the independent clauses $C_0$ and $C_1$ simultaneously, followed by the intersecting third clause $C_2$.

\myparagraph{Algorithm} Graph coloring is a well-studied problem. The optimal solution is the one that requires the least colors. The NP-hard nature of the problem makes this task challenging. Brute-force solutions are intractable as the search space increases exponentially. We settle for a heuristical algorithm, DSatur, with quadratic complexity and quality results. Based on DSatur~\cite{dsatur} greedy coloring approach, Algorithm~\ref{alg:clause_coloring} shows its implementation using the example in Figure~\ref{fig:woptimizer-overview}.

\begin{algorithm}[t]
    \caption{Pseudocode for color shuttling}
    \fontsize{8}{9}\selectfont
    \label{alg:color_shuttling}
    \begin{algorithmic}
        \Require $S \gets$ Ordered list of atoms in the current color zone, $F \gets$ ordered list of atoms in the next color zone
        \Ensure $R \gets$ List of shuttling instructions
        \State $R \gets []$
        \For{$a \in S$}
            \If{$a \in F$}
                \State $\texttt{transfer\_to\_aod(a)}$ \Comment{\texttt{Used in next color}}
            \Else
                \State $\texttt{transfer\_to\_slm(a)}$ \Comment{\texttt{Unused atom}}
            \EndIf
        \EndFor
        \While{there is an atom $a_i \in F$ that is not yet scheduled}
            \State $W \gets \{a_i\}$ \Comment{\texttt{Current shuttle set}}
            \For{$a_j \in F$ and $a_j$ not yet scheduled}
                \If{Order between $a_i$ and $a_j$ is same in $S$ and $F$}
                    \State $\texttt{W.add(}a_j\texttt{)}$ \Comment{\texttt{shuttle in parallel}}
                \EndIf
            \EndFor
            \State $\texttt{R.add(create\_shuttle(W))}$
        \EndWhile
        \State \Return $R$
    \end{algorithmic}
\end{algorithm}

\subsection{Color Shuttling}

\myparagraph{Challenge} QAOA circuits require arbitrary 2-qubit connections; connections that lie on two unconnected qubits are normally implemented in superconducting with $SWAP$ gates executed through 3 $CNOT$ gates~\cite{NielsenChuang2010}, worsening the fidelity loss. In FPQAs, the non-physical connection of qubits and the availability of AOD traps enable qubit reconfiguration. This hardware feature motivates the replacement of logical routing operations with physical ones using qubit shuttling.

\myparagraph{Key idea} FPQAs' qubit reconfiguration must follow specific constraints that prevent arbitrary atom movement, making the formulation of a strategy for general quantum circuits tough. However, the clause coloring stage provides a clear conceptual strategy, as depicted in Figure~\ref{fig:woptimizer-overview}.

Each color group requires two AOD columns and one AOD row, and they are executed sequentially. Color groups are set diagonally to avoid AOD constraints described in Section ~\ref{section:wqasm}. After completing a color, the atoms for the upcoming color are transferred to AOD traps and shuttled to their next locations.

Parallel shuttling is a simple task; as long as the order between atoms in a color zone is kept in the AOD row, they can be shuttled in parallel without violating any AOD constraint. In our example, after executing the first color, the order of the atoms in the AOD row will be $x_2 > x_4 > x_5$. However, the new expected order is $x_4 > x_2 > x_5$ in the next color zone. This requires a two-step shuttle: first, transferring $x_4$ and $x_5$ together, followed by $x_2$, to proceed with the next zone.

\myparagraph{Algorithm} The Color shuttling stage processes atom orders within and between color zones, transferring atoms to the next diagonal zones, as described previously and detailed in Algorithm~\ref{alg:color_shuttling}. The implementation of the shuttling instruction is not shown on the pseudocode, which is trivial if the order of clauses within a color is fixed before compilation time.

\subsection{3-qubit Gate Compression}

\myparagraph{Challenge} Using multi-qubit gates is challenging because most gate compilers focus on 2-qubit gate decompositions. Moreover, higher-order gates' fidelity is less than lower-order ones. We must ensure that enough lower-order gates are compressed in the circuit to offset the worse fidelity of the higher-order gate. Additionally, research mainly targets compilation to $CNOT$ gates. However, FPQAs use $C_n Z$ gates.

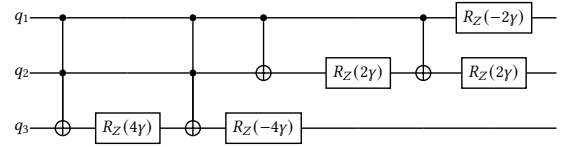
\begin{figure}
    \centering
    \begin{adjustbox}{width=0.9\linewidth}
    \begin{quantikz}
        & q_1 & \ctrl{2} & \qw & \ctrl{2} & \ctrl{1} & \qw & \ctrl{1} & \gate{R_Z(-2\gamma)} & \qw \\
        & q_2 & \ctrl{1} & \qw & \ctrl{1} & \targ{} & \gate{R_Z(2\gamma)} & \targ{} & \gate{R_Z(2\gamma)} & \qw \\
        & q_3 & \targ{} & \gate{R_Z(4\gamma)} & \targ{} & \gate{R_Z(-4\gamma)} & \qw & \qw & \qw & \qw \\
    \end{quantikz}
    \end{adjustbox}
    \vspace{-3mm}
    \caption{Implementation of the first clause $(\neg x_0 \lor \neg x_1 \lor \neg x_2)$ from the example in Figure~\ref{fig:woptimizer-overview} (\S~\ref{section:woptimizer}). While the CCNOT part implements the terms $x_0 x_2$, $x_1 x_2$ and $x_0 x_1 x_2$, the rest of the circuit computes the missing single variable terms and the quadratic term $x_0, x_1$.}
    \label{fig:ccnot_compression}
\end{figure}

\myparagraph{Key idea} MAX-3SAT requires 3-qubit interactions. The key idea is to implement each clause with $CCNOT$ gates instead of solely $CNOT$ gates. However, $CCNOT$ gates are not enough to implement all the terms in the clause, as shown in Figure~\ref{fig:ccnot_compression}, due to the symmetric behavior of the $CCNOT$ gate regarding its control qubits. The interactions between the control qubits must be implemented with a separate $CNOT$ ladder. A 3-qubit gate is, thus, implemented with 2 $CCNOT$ and 2 $CNOT$ gates, a more efficient approach compared to using 8 $CNOT$ gates.

\myparagraph{Algorithm} Gate compression largely depends on the difference of fidelity parameters of $CZ$ and $CCZ$ gates on the target hardware. The compression stage first determines whether using the compression is beneficial; if so, it identifies the appropriate sub-circuit for each clause in the current color zone, which depends on the number of negative/positive literals in the clause. For clauses with mixed literals, the control bits in the $CCNOT$ gate are set to zero with single-qubit rotation gates. Other cases can be corrected by adjusting the signs of the angles as in Figure~\ref{fig:ccnot_compression}. The algorithm keeps track of each clause's control qubits throughout the execution of the formula. This helps us to calculate the term between the control qubits with a single 2-qubit interaction instead of repeated applications. The actual implementation of the sub-circuit can be seen in Figure~\ref{fig:woptimizer-overview}. Firstly, the atoms in a clause are positioned in a triangular layout employing a global Rydberg pulse for the $CCNOT$ section, and then the control qubits are repositioned to facilitate or avoid additional interactions. Afterward, the control qubits are shuttled apart from the target qubit to implement the missing quadratic interaction. An additional shuttle can be applied to control qubits to prevent unnecessary quadratic terms.

\begin{figure*}[t]
    \centering
    \includegraphics[width=0.9\textwidth]{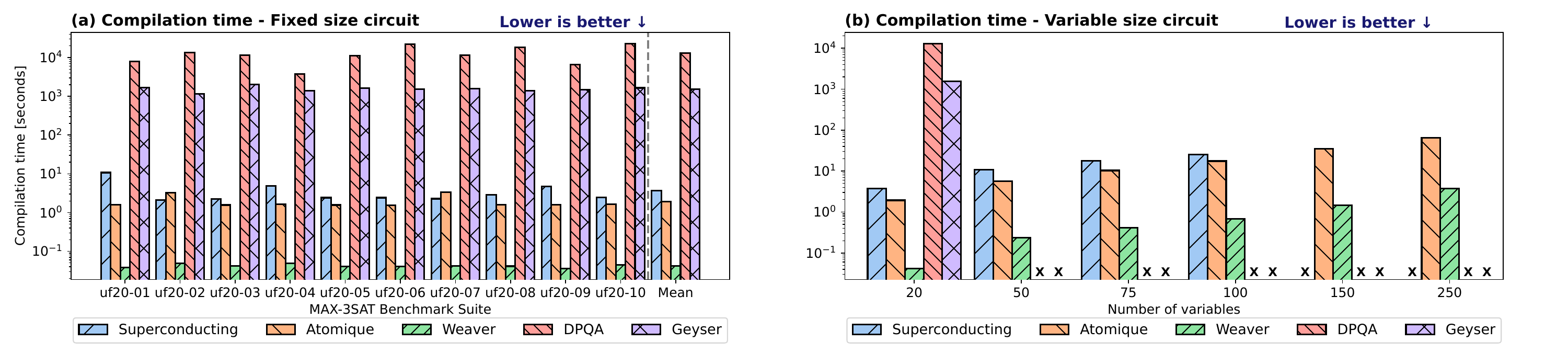}
    \caption{Compilation time (\S~\ref{sec:evaluation:compile-time}). {\bf (a)} Compilation time for fixed-size circuits of 20 variables. {\bf (b)} Compilation time for variable size circuits. Geyser and DPQA timed out above 20 variables. Superconducting was run with up to 100 variables, limited by the 127-qubit available backends.}
    \label{fig:evaluation:compilation_time}
    \vspace{-2mm}
\end{figure*}

\subsection{Complexity Analysis}
The complexity of \projectname's \woptimizer is bound by the clause coloring procedure, which follows the $O(N^2)$ complexity of DSatur \cite{dsatur}.
The color shuttling step loops through the traps in the AOD from end to start, checks if the atom is used for the next color group, and schedules a corresponding shuttling instruction. If a previous shuttling operation exists and it does not collide with the current one, they are then merged together and executed in parallel. The first input to Algorithm \ref{alg:color_shuttling}, i.e., the current positions of the atoms, is calculated by the previous coloring step. Since we loop the atoms through the AOD traps, it is guaranteed that the order is preserved, which is a hard constraint for Algorithm \ref{alg:color_shuttling}. Therefore, the color-shuttling procedure requires no additional sorting. 
As the coloring procedure precomputes execution sites (the second input for Algorithm \ref{alg:color_shuttling}) and the colors are sorted once with $O(N \log N)$ complexity, shuttling a single color takes $O(N)$, resulting in $O(N^2)$ complexity for at most $O(N)$ colors. 3-qubit gate compression for a single color takes $O(N)$ time, as all the individual shuttling instructions, Raman, and Rydberg pulses can be implemented in constant time. Similar to color shuttling, since there are at most $O(N)$ colors, the overall complexity of 3-qubit gate compression also becomes $O(N^2)$. Thus, \woptimizer's complexity is determined by the $O(N^2)$ complexity of the coloring step for $N$ variables/qubits/atoms.
\begin{figure}
    \centering
    \includegraphics[width=0.85\linewidth]{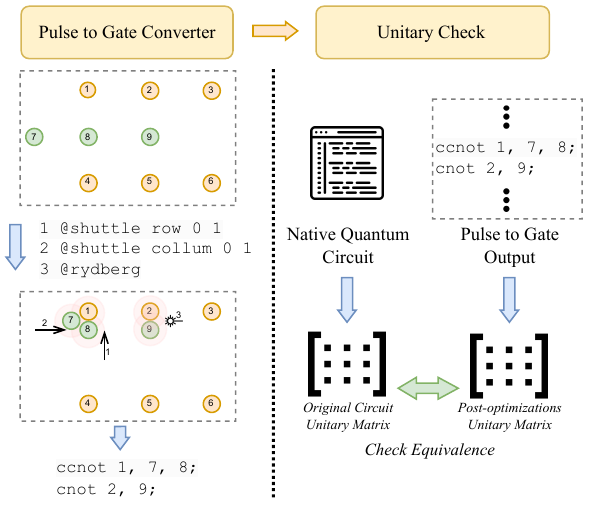}
    \vspace{-3mm}
    \caption{Example workflow of the \wchecker (\S~\ref{section:wchecker}).
     \wchecker comprises a pulse-to-gate converter and a unitary checker. Initially, three pulse instructions are translated into their respective gate instructions. A \texttt{@shuttle} instruction moves the row with qubits $7$, $8$, and $9$ up, and a second instruction moves the column with qubit $7$ to the right. A Rydberg pulse applies $ccnot$ to $1$, $7$, $8$, and $cnot$ to $2$, $9$. Then, a native quantum circuit is converted to a unitary matrix with gate instructions for the comparison of two unitary gates.}
    \label{fig:wchecker}
\end{figure}

\begin{figure*}[t]
    \centering
    \includegraphics[width=0.85\textwidth]{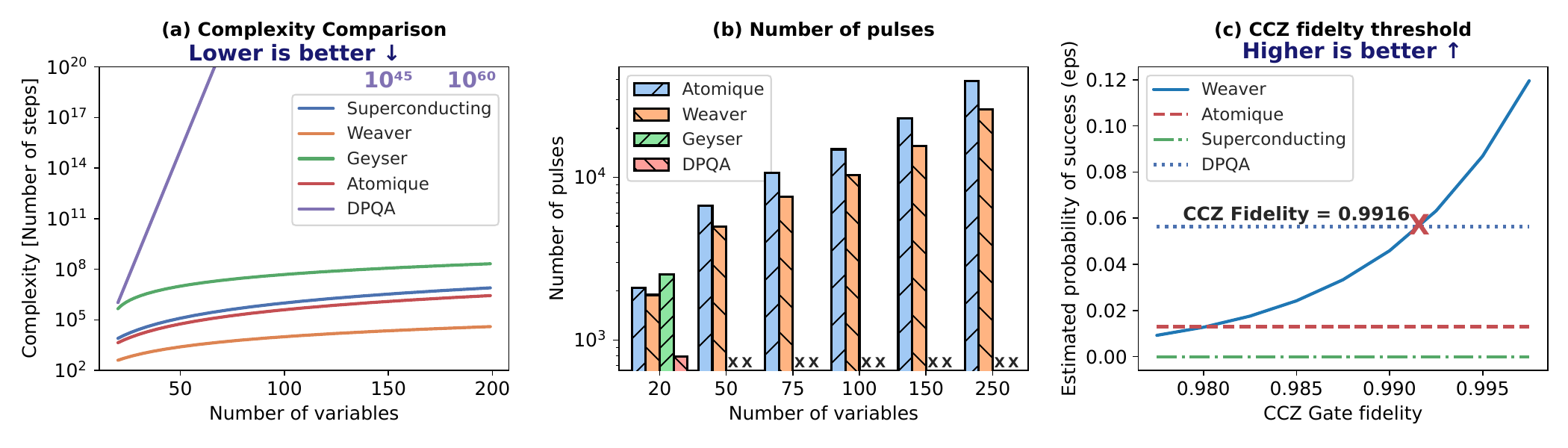}
    \vspace{-1mm}
    \caption{Performance comparison between \projectname and baselines (\S~\ref{section:evaluation}). {{\bf (a)} Visual representation of compilation complexity of each system, also represented on Table \ref{tab:complexity}. {\bf (b)} Comparing the average number of gates in the solutions generated by each system. {\bf (c)} Threshold complexity of CCZ gate where Weaver's solution's fidelity surpasses all baselines.}}
    \label{fig:evaluation:analysis_plot}
    \vspace{-2mm}
\end{figure*}

\section{\wchecker Equivalence Checker}
\label{section:wchecker}
The \wchecker checks the functional equivalence between the nativized and the FPQA-optimized circuits. To achieve this, it checks the \wqasm file, which contains hardware-level instructions for each logical gate and hardware-specific Rydberg and Raman pulses for FPQAs, to see if these pulse instructions can mirror their logical gate equivalent.
To maintain logical circuit integrity, (global or local) Raman pulses can be verified by comparing their pulse angles and addresses to their logical specifications, as outlined in \S~\ref{section:wqasm}. Conversely, Rydberg pulses are more challenging to check, given that we need to know the positions of the atoms inside the FPQA beforehand. As such, \wchecker must simulate the atom movements (atom transfers and shuttle instructions) before each Rydberg pulse instruction. The SLM and AOD initialization instructions at the start of a \wqasm file determine the number of atoms and their starting positions in the system.

Figure~\ref{fig:wchecker} shows \wchecker in action, specifically the verification of a logical $CCZ$ gate between the qubits 1, 7, and 8. \wchecker sequentially simulates the two shuttle instructions and then checks each atom pair according to the new positions of the atoms. For correct compilation, \wchecker has to verify that \textbf{(1)} the atoms 1, 7, and 8 are interacting with each other, \textbf{(2)} they are in equal distance to each other, \textbf{(3)} no other atoms are currently interacting with any other atoms.

The complexity of \wchecker is $O(N^2M)$ steps for a circuit with $N$ qubits and $M$ instructions, bound by Rydberg pulses, which require checking each qubit pair for interactions, causing quadratic scaling with qubits. Each remaining instruction can be checked in at most $O(N)$ time. For example, validating a shuttling instruction involves looping through AOD columns/rows to ensure no crossover with neighbors. Likewise, single-qubit gates, i.e., Raman pulses, can be checked in $O(1)$ time.
\section{Implementation}
\label{section:implementation}

We implement \projectname on top of Qiskit 1.0.2~\cite{qiskit2024} and OpenQASM 3.0~\cite{openqasm3}, and use PySAT 3.2.0~\cite{imms-sat18} for the \woptimizer module. The native gate synthesis in Figure~\ref{fig:weaver-overview} is done by the Qiskit compiler by setting the appropriate basis gate set, $B=\{U3, CZ\}$.
To retarget a \wqasm file for a superconducting device, we use the Qiskit compiler \cite{qiskit-transpiler}.

\woptimizer uses PySAT to handle MAX-3SAT formulas and represents the FPQA device as a class with adjustable hardware parameters for compilation. Since FPQAs are still emerging, \projectname aims to remain hardware-agnostic, supporting flexible SLM trap layouts and varying numbers of AOD rows and columns. A key simplification is that it assumes only digital computation; for example, if three atoms are within range and a Rydberg pulse is applied, it treats the operation as a $CCZ$ gate, which is accurate only if the atoms are equidistant. If not, the operation could correspond to a different multi-qubit gate.

\wchecker begins by parsing the \wqasm file using Qiskit's OpenQASM passes, then processes each FPQA annotation in a visitor pattern to reconstruct the FPQA class created by \projectname during compilation. Each FPQA instruction shares a basic interface for validating the operation and generating the equivalent \wqasm instruction with annotations. As detailed in \S~\ref{section:wchecker}, \wchecker compares the FPQA annotations against the logical circuit in the output \wqasm file.

\section{Evaluation}
\label{section:evaluation}
We evaluate \projectname across three dimensions: (a) compilation time (\S~\ref{sec:evaluation:compile-time}), (b) execution time (\S~\ref{sec:evaluation:execution_time}), and (c) fidelity (\S\ref{sec:evaluation:fidelity}).

\subsection{Experimental Methodology}
\label{subsec:experimental_met}

\myparagraph{Experimental setup} We run \projectname and all the baselines on a server with a 64-core AMD EPYC 7713P processor and 1TB of DDR4 memory. We use \textit{IBM Washington} \cite{ibmFakeWashingtonQuantum} as our superconducting backend model.

\myparagraph{Evaluation metrics}
We evaluate \textbf{(1)} compilation time in seconds, \textbf{(2)} execution time in seconds, and the number of pulses generated, and \textbf{(3)} fidelity as the Estimated Probability of Success (EPS) \cite{schmid2024computational}.

\myparagraph{Benchmarks} We use the MAX-3SAT formulas from the SATLIB benchmark \cite{satlib}. These formulas define statements with varying numbers of clauses and variables. More clauses lead to longer quantum circuits, while the number of variables corresponds to the number of qubits.

\myparagraph{Experimental methodology}
We run \textbf{(1)} 10 different MAX-3SAT problems with the fixed size of 20 variables and \textbf{(2)} with an increasing number of variables, where each data point is the average of the 10 SAT problems of that size.

\myparagraph{Framework and configuration} We use Qiskit version 1.0.2 with FPQA hardware parameters from \cite{Schmid_2024} and \cite{evered_high-fidelity_2023} based on Rubidium atoms. We set a timeout of 20 hours for the evaluated compilers to find a solution.

\myparagraph{Baselines} We compare \projectname to state-of-the-art FPQA compilers: \textit{Geyser} \cite{patel2022geyser}, \textit{Atomique} \cite{wang_atomique_2024}, and \textit{DPQA} \cite{tan_compiling_2024}. For superconducting, we use \textit{Qiskit} \cite{qiskit-transpiler} compiler.

\begin{figure*}[t]
    \centering
    \includegraphics[width=0.9\textwidth]{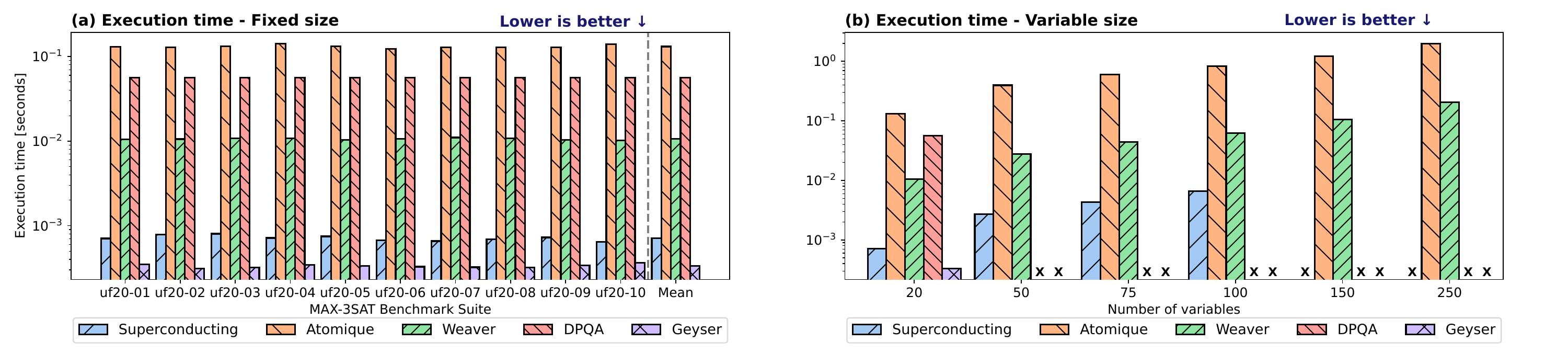}
    \caption{Execution time (\S~\ref{sec:evaluation:execution_time}) {{\bf (a)} Execution time for fixed-size circuits of 20 variables. {\bf (b)} Execution time for increasing benchmark sizes. Geyser and DPQA timed out above 20 variables. Superconducting was run with up to 100 variables, limited by the 127-qubit available backends.}}
    \label{fig:evaluation:execution_time}
\end{figure*}

\subsection{Compilation Time}
\label{sec:evaluation:compile-time}
{\bf RQ1:} \textit{What is \projectname's performance w.r.t. compilation time, compared to the baselines?}

\myparagraph{Hypothesis}
We measure the end-to-end compilation time of \projectname and the baselines using the \texttt{time} library. 
\projectname should achieve lower compilation times than the baselines due to lower computational complexity, as shown in Table \ref{tab:complexity}. \projectname scales quadratically as the number of variables increases, while the baselines scale cubically or with the number of quantum operations, which are generally significantly more than the number of variables.

\myparagraph{Results}
Figure \ref{fig:evaluation:compilation_time} (a) shows the compilation time for 10 benchmarks with a size of 20 variables. Figure \ref{fig:evaluation:compilation_time} (b) shows the results of increasing the circuit size of the MAX3SAT problems from 20 up to 250 variables. \projectname consistently finds solutions $5.7x10^3 \times$ faster than all other systems. \textit{Geyser} and \textit{DPQA} are the most time-intensive systems, on average $1.5x10^3 \times$ slower than Superconducting, \textit{Atomique} and \textit{Weaver}.

\myparagraph{Analysis}
Figure \ref{fig:evaluation:analysis_plot} (a) shows the compilation complexity comparison as the number of steps. Geyser's complexity is based on the number of quantum operations on the benchmark circuit. For this, 6 circuits with sizes from 20 to 250 were used to find a fitting function to express Geyser's complexity based on the number of variables.
Notably, \projectname exhibits the lowest complexity as the number of variables increases.

\myparagraph{RQ1 takeaway} \projectname's heuristics achieve faster compilation times than the baselines, up to $24 \times$ lower than Atomique; up to $10^5 \times$ lower than DPQA and $10^4 \times$ lower, on average.

\subsection{Execution Time}
\label{sec:evaluation:execution_time}

\begin{table}
    \centering
    \caption{Compilation complexity comparison (\S~\ref{sec:evaluation:compile-time}). {\em $N$ is the number of benchmark variables, and $K$ is the number of quantum circuit operations (generally, $K \gg N$). The complexity of \textit{Qiskit} and \textit{Atomique} stem from Sabre \cite{gushu2019tackling}.
    }}
    \vspace{4pt}
    \fontsize{8}{9}\selectfont
        \begin{tabular}{c|c}
            \textbf{Compiler} & \textbf{Computational complexity} \\
            \hline \hline
            Qiskit \cite{qiskit-transpiler} & $O(N^3)$ \\
            \hline                      
            Atomique \cite{wang2024atomique} & $O(N^3)$ \\
            \hline
            Geyser \cite{patel2022geyser} & $O(K^2)$ \\
            \hline  
            DPQA \cite{tan_compiling_2024} & $O(2^K)$ \\
            \hline
            Weaver & $O(N^2)$ \\
            \hline
        \end{tabular}
    
    \label{tab:complexity}
\end{table}

{\bf RQ2:} \textit{What is \projectname's performance w.r.t. execution time, compared to the baselines?}

\myparagraph{Hypothesis} We measure how long the quantum circuit runs on a quantum device by adding the times of each pulse and shuttling operation, considering the maximum movement speed. Typically, longer circuit duration indicates a higher chance for decoherence errors \cite{gushu2019tackling}; hence, lower is better.  

Minimizing execution times is challenging due to the NP-hard nature of finding optimal solutions at various stages in compilation \cite{cowtan2019qubit} and \projectname aims to balance compilation time and execution time. We expect circuits compiled with \projectname to have lower execution times since \projectname increases parallelization, as detailed in \S~\ref{section:woptimizer:clause_coloring}.

\myparagraph{Results}
Figure \ref{fig:evaluation:execution_time} (a) shows the execution time of \projectname's solutions compared to the baselines. \projectname consistently achieves $8.2 \times$ better execution times than \textit{Atomique} and \textit{DPQA}. Superconducting (\textit{Qiskit}) has faster quantum gate times than FPQAs, which explains the results.

Although Geyser incurs the lowest execution times, it times out for problems larger than 20 variables, as shown in Figure \ref{fig:evaluation:execution_time} (b).

\myparagraph{Analysis}
Figure \ref{fig:evaluation:analysis_plot} (b) presents the mean number of laser pulses. FPQA pulses for atom shuttling are time-consuming, given that the movement must be relatively slow to avoid losing the atom or inducing noise. Geyser does not use atom movement in its solutions, which explains its fast solutions while showing many pulses. DPQA applies the most atom movement of all systems, which explains the low number of pulses but longer execution time.

\myparagraph{RQ2 takeaway} \projectname's parallelization strategy and the low number of pulses reduce execution times by $12.4\times$ in the best case and $4.4\times$ on average, compared to the baselines.

\subsection{Fidelity}
\label{sec:evaluation:fidelity}

\begin{figure*}[t]
    \centering
    \includegraphics[width=0.9\textwidth]{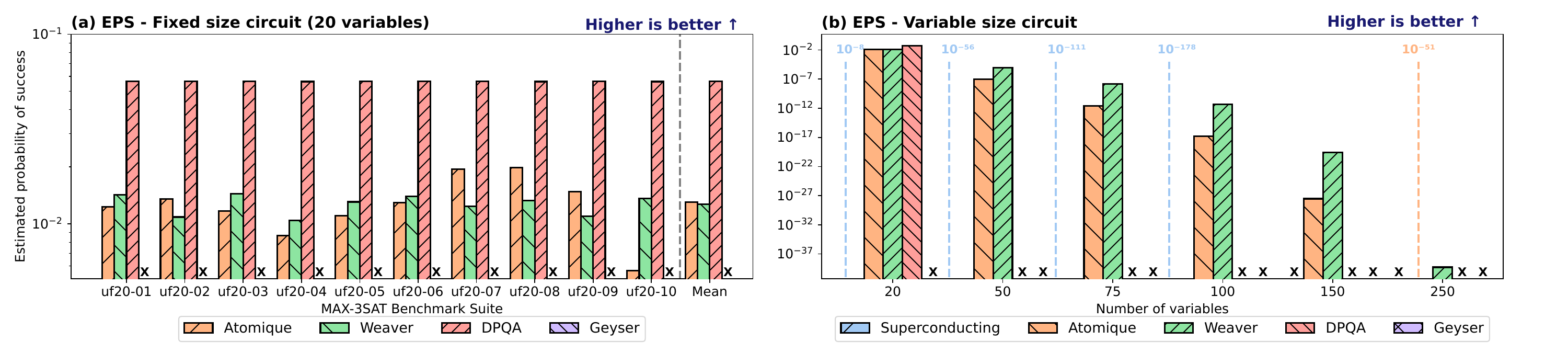}
    \caption{Fidelity as EPS (\S~\ref{sec:evaluation:fidelity}). {{\bf (a)} Execution time for fixed-size circuits of 20 variables. {\bf (b)} Execution time for increasing benchmark sizes. Geyser was not considered since the block approximation step causes EPS computation to be unfair. DPQA timed out above 20 variables.}}
    \label{fig:evaluation:fidelity}
\end{figure*}

{\bf RQ3:} \textit{What is \projectname's performance w.r.t. EPS, compared to the baselines?}

 \myparagraph{Hypothesis} EPS measures the likelihood that a circuit runs correctly in one execution, calculated by accumulating the errors of each pulse operation. \projectname aims to improve EPS compared to the baselines, using heuristics to tackle this NP-hard problem and balancing between compilation time and the EPS of the solutions found.

\myparagraph{Results}
Figure \ref{fig:evaluation:fidelity} shows the EPS of \projectname solutions compared to the baselines. \projectname consistently achieves better EPS than other baselines except for \textit{DPQA}, which finds a better solution for 20 variables benchmarks. Increasing the benchmark sizes increases \projectname's improvement on EPS over its closest competitor, \textit{Atomique}.

\myparagraph{Analysis}
\projectname's approach leverages 3-qubit gates (CCZ) despite their current susceptibility to higher error rates. This distinguishes \projectname from most baseline approaches, which avoid 3-qubit gates to minimize error. Figure \ref{fig:evaluation:analysis_plot} (c) illustrates the threshold of the CCZ gate fidelity required for \projectname to outperform all pictured baselines on a 20-variable benchmark. The threshold of 0.9916 is only a $1.2\%$ improvement over the currently used CCZ error of $0.98$.

\myparagraph{\bf RQ3 takeaway} \projectname achieves on average $10\%$ improvement in EPS when compared to Atomique, evaluated on 20 variable benchmarks, while for a larger benchmark of 150 variables, the improvement is on the order of $10^8 \times$.
\section{Related Work}
\label{section:related_work}

\myparagraph{Hardware-independent compilers and optimizations}
This is an active area of research, and the following non-exhaustive list features universally applicable compiler optimizations (even when framed for superconducting QPUs) \cite{sivarajah2020t, li2022paulihedral, das2021jigsaw, tannu2019mitigating, bravyi2021mitigating, maciejewski2020mitigation, yunong2019optimized, gokhale2020optimized, das2023the, xu2023synthesizing, patel2022quest, li2024qutracer, seif2024suppressing}. Notably, this work is orthogonal to \projectname{} and can be an additional compiler stage. For instance, the optimizations of \textit{Paulihedral} \cite{li2022paulihedral} can be applied before \projectname{}, while readout error mitigation \cite{das2021jigsaw, tannu2019mitigating, seif2024suppressing} is applicable as a post-processing step.

\myparagraph{Optimizations for superconducting QPUs}
Superconducting QPUs are extensively studied and their optimizations and can be categorized as follows: (1) qubit mapping and routing \cite{gushu2019tackling, prakash2019noise, swamit2019not, chi2021time, molavi2022qubit, Zulehner2019anefficient, wille2019mapping, siraichi2018qubit, patel2021qraft, patel2020veritas, jin2024exploiting, tannu2019ensemble}, (2) instruction/pulse scheduling \cite{das2021adapt, tripath2022suppression, smoth2022timestitch, murali2020software, litteken2023dancing, chen2024one}, and (3) error suppression/mitigation \cite{patel2020disq, Maurya2023scaling, das2021jigsaw, tannu2019mitigating}. However, such work does not apply to FPQAs due to its specificity to superconducting's limitations (i.e., limited connectivity and short coherence times), while the unique features of FPQAs as described in \S~\ref{sec:quantum_architectures} directly address these limitations.

\myparagraph{Compilers for neutral atoms/FPQAs}
Compiler frameworks and optimizations for neutral atoms are an emerging area of research. Implementing these compilers to leverage the capabilities of neutral atoms, as discussed in \S~\ref{sec:quantum_architectures}, presents a set of challenging problems. \textbf{(1)} While qubit shuttling is utilized in several studies \cite{brandhofer2023optimal, tan2022qubit, nottingham2024circuitdecompositionsschedulingneutral, wang2024atomique, wang2024qpilot}, not all neutral atom compilers take advantage of atom shuttling; some systems opt for a fixed atom grid and rely on SWAP operations \cite{patel2022geyser, 10082942qtetris, baker2021exploiting}.
 \textbf{(2)} Supporting 2+ qubit gates is another challenging capability in neutral atom architectures, primarily due to the difficulties in synthesizing quantum circuits to utilize gates with more than two qubits. Currently, only Geyser \cite{patel2022geyser} achieves this by approximating blocks of the original circuit to a template sequence composed of unitary and Toffoli gates. Other compilers can map and route circuits with multi-qubit gates if they are included in the quantum circuit in the first place \cite{baker2021exploiting, 10082942qtetris, schmid2024computational}. 
In contrast, \projectname{} leverages all FPQA capabilities, including both qubit shuttling and three-qubit gates, making it the first compiler, to our knowledge, to do so.

\myparagraph{Retargetable quantum compilers}
Research on retargetable compilers is growing since different quantum technologies provide different advantages that are beneficial to different applications. $t\ket{ket}$ \cite{sivarajah2020t} is able to target different quantum superconducting providers (IBM Q, Rigetti, ProjectQ, etc.) while sharing circuit optimizations before the circuit is compiled to a certain vendor's API. On the other hand, XACC \cite{mccaskey2019xaccsystemlevelsoftwareinfrastructure} provides support for different quantum technologies and tries to share optimizations when possible. This approach may not be optimal since applying architecture-specific optimizations can benefit the quantum circuit performance. \projectname{} targets superconducting and neutral atoms technology through a common input QASM file but applies distinct optimizations to each technology.

\myparagraph{Domain-specific optimizations}
There exist optimizations leveraging the distinctive features of particular algorithms and/or circuit structures to improve fidelity beyond the capabilities of general-purpose compilers \cite{mahabubul2020circuit, lao20222qan, gokhale2019partial, stein2022eqc, hao2023enabling, tuysuz2023classical, ravi2022cafqa, ravi2023navigating, dangwal2023varsaw, jin2024tetris, wang2024redqaoa, anagolum2024elivagar, Ayanzadeh2023frozenQbits}. However, the listed work above either (1) is orthogonal, i.e., can be applied as a pre/post-processing step \cite{Ayanzadeh2023frozenQbits, stein2022eqc, hao2023enabling} or targets a different domain (\cite{lao20222qan}), (2) is specific to superconducting QPUs (\cite{mahabubul2020circuit}) or (3), focuses on compiler performance w.r.t. runtime only (\cite{gokhale2019partial}).

\section{Conclusion}
\label{section:conclusion}
While superconducting devices are the leading quantum technology, they face scalability issues, which are addressed by the emerging and promising FPQA technology. To this end, we presented \projectname, a retargetable compiler framework 
that retargets existing superconducting code to FPQAs while leveraging their distinct capabilities, namely the dynamic qubit connectivity, higher-order gates, and high gate parallelism. \projectname compiles circuits $\approx10^3\times$ faster, while the compiled circuits have $4.4\times$  lower execution times and up to $10^5\times$ higher execution fidelity.
\section*{Acknowledgements}

We would like to thank the anonymous reviewers for their valuable insights. This work was supported by the Bavarian State Ministry of Science and the Arts with funds from the Hightech Agenda Bayern Plus, as part of the Munich Quantum Valley (MQV) initiative (6090181).
\appendix
\section{Artifact Appendix}

\subsection{Abstract}
Our artifacts include Weaver's framework, focusing on Weaver's proposed FPQA optimization and all the other baseline FPQA compilers for comparison. These include Geyser \cite{patel2022geyser}, Atomique \cite{wang2024atomique} and DPQA \cite{tan_compiling_2024}. The artifact also includes code to run with Qiskit \cite{qiskit-transpiler} for a Superconducting compiler baseline. Weaver's compilation procedure takes advantage of the 3 variable sets of the MAX-3SAT problems as explained in \ref{section:woptimizer}. This appendix provides the necessary information to install all the required Python libraries and reproduce the experiments presented in the paper. Our artifacts demonstrate Weaver's ability to achieve $10^3\times$ faster execution time than baseline approaches, as well as $4.4\times$ better compilation time and $10\%$ on average better fidelity.

\subsection{Artifact Check-list (Meta-information)}
\begin{itemize}
    \item \textbf{Program:} Weaver FPQA optimization's source code. The source codes for Geyser, Atomique, and DPQA baseline compilers are included. The source code for the Qiskit compiler can be installed as a Python library.
    \item \textbf{Data set:} The MAX-3SAT problems used are included in the artifact and were taken from the SATLIB benchmark \cite{satlib}.
    \item \textbf{Hardware:} The artifact does not require any specific hardware or features.
    \item \textbf{Software:} Weaver and the other baseline compilers were tested on Python 3.11.2.
    \item \textbf{Execution:} The artifact needs to execute each of the baseline compilers for the given benchmark circuits. Weaver, Qiskit, and Atomique should each take no more than 1 hour to run all the benchmarks. However, Geyser, especially DPQA, might require up to 10 hours each to run all benchmarks. 
    \item \textbf{Metrics:} Execution time, compilation time, and result fidelity are the main three metrics that were measured; additionally, to complement each of these metrics, we also measure algorithm complexity, number of pulses, and the CCZ fidelity threshold.
    \item \textbf{Output:} The artifact should output the 4 figures presented on the paper, which should be similar to the ones shown, with some negligible variation.
    \item \textbf{Experiments:} Preparing the experiments should take no more than 10 minutes; this involves downloading the repo and setting up the Python environment. The whole experiment can be left running in the background. However, it may take up to 24 hours to complete.
    \item \textbf{Publicly available:} Yes.
    \item \textbf{License:} MIT License. Weaver doesn't use any external license.
    \item \textbf{Archive:} Weaver's artifact is archived at: \url{https://doi.org/10.5281/zenodo.14084047} \cite{kirmemis_2024_14084047}
\end{itemize}

\subsection{Description}

\subsubsection{Artifact}
All the artifact's source code can be found at \url{https://github.com/TUM-DSE/weaver}.

\subsubsection{Benchmarks}
The benchmarks used are taken from a collection of MAX-3SAT problem formulations used in MAX-3SAT solvers competitions. The problem formulations used are of the sizes 20, 50, 75, 100, 150 and 250. Each problem size has 10 variants with different combinations of variables.

\subsection{Installation}
All the required libraries are contained in the file \path{pyproject.toml}. To install the dependencies, it is suggested to create a virtual environment with any virtual environment Python framework or follow these steps:

\begin{enumerate}
    \item \texttt{python -m venv .venv}
    \item \texttt{pip install pdm}
    \item \texttt{pdm install}
\end{enumerate}

After \texttt{pdm} installs all the packages and dependencies, you can move to execute the \texttt{run} script.

\subsubsection{Experiment Workflow}
To reproduce the plots presented in the paper, one simply needs to run: \texttt{python run.py} on the main folder.
The script will execute 7 scripts and finally plot 4 figures. The scripts are the following:

\begin{enumerate}
    \item \textbf{MAX-3SAT to quantum circuit $[\approx10\ minutes]$:} \\ Transpiles MAX-3SAT instances to QAOA quantum circuits with Qiskit and basis gates \textit{RX, RZ, X, Y, Z, H, ID, CZ}. Since most baseline compilers get a quantum circuit as input, the transpilation from MAX-3SAT instances to the quantum circuit is done here to save time and keep experiment consistency.
    \item \textbf{Atomique $[\approx30\ minutes]$:} Runs Atomique compiler on the previously transpiled quantum circuits. Atomique compiles quantum circuits of all sizes (20, 50, 75, 100, 150, 250) and 10 variant circuits for each size.
    \item \textbf{Superconducting (Qiskit) $[\approx20\ minutes]$:} Runs \\ Qiskit transpiler on the previously transpiled circuits. This script runs the compiler for sizes (20, 50, 75, and 100). It doesn't run for larger circuits since the targetted backend (\textit{Washington}) is of size 127 qubits. Again, each size runs for 10 variants.
    \item \textbf{Geyser $[\approx 7\ hours]$:} Runs Geyser compiler for 10 different circuits of size 20 variables. Running Geyser on larger circuits takes longer than the defined timeout compilation time (20 hours).
    \item \textbf{Weaver $[\approx 20\ minutes]$:} Runs Weaver compilation procedure. Similarly to Atomique, Weaver also runs for all sizes of the benchmarks and all variants.
    \item \textbf{Quantum circuit to DPQA format $[\approx 2\ minutes]$:} Converts a quantum circuit to the format required by the DPQA compiler. The format is a \texttt{.json} file with sets of two-qubit gates. Expected duration \~10 minutes.
    \item \textbf{DPQA $[\approx 15\ hours]$:} Runs the DPQA compiler. It runs 10 variants of benchmarks of size 20. For the same reason as Geyser, we didn't run the compiler on larger benchmarks.
    \item \textbf{Plotting $[\approx 1\ minute]$:} Finally, the script plots the 4 figures presented on the paper based on the results of the previous compilations: Figure \ref{fig:evaluation:compilation_time}, Figure \ref{fig:evaluation:analysis_plot}, Figure \ref{fig:evaluation:execution_time}, and Figure \ref{fig:evaluation:fidelity}. The complexity comparison Figure \ref{fig:evaluation:analysis_plot} (b) and CCZ fidelity threshold Figure \ref{fig:evaluation:analysis_plot} (c) plots are the only ones that plot fixed lines and values pre-calculated.
\end{enumerate}

\subsection{Evaluation and Expected Results}
Throughout the experiment, the script will output information about its progress; this is a hint that the script is still running and hasn't halted, also considering how long each step of the script takes. The output plots may have some variance when compared to the plots presented on the paper, however this variance is not expected to surpass $5\%$.

\subsubsection{Methodology}
Submission, reviewing, and badging methodology:
\begin{itemize}
    \item \url{http://cTuning.org/ae/submission-20190109.html}
    \item \url{http://cTuning.org/ae/reviewing-20190109.html}
    \item \url{https://www.acm.org/publications/policies/artifact-review-badging}
\end{itemize}

\balance

\bibliographystyle{ACM-Reference-Format}
\bibliography{references}

\end{document}